\documentclass[fleqn,10pt]{wlscirep}

\usepackage{amsmath} 
\usepackage{amssymb} 
\usepackage{amsfonts}
\usepackage{endfloat}
\graphicspath{{eps/}}

\renewcommand\refname{References}

\newcounter{lastnote}

\newcommand{\vc}[1]{{\bf #1}}

\newcommand{\ma}[1]{{\bf #1}}

\title{Steering Macro-Scale Network Community Structure by Micro-Scale Features}

\author[1,2]{Dimitri Van De Ville}
\affil[1]{Institute of Bioengineering, Center for Neuroprosthetics, Ecole Polytechnique F\'ed\'erale de Lausanne, Switzerland}
\affil[2]{Department of Radiology and Medical Informatics, University of Geneva, Switzerland}
\affil[*]{dimitri.vandeville@epfl.ch}

\begin{abstract}
Network science plays an increasingly important role to model complex data in many scientific disciplines. One notable feature of network organization is community structure, which refers to clusters of tightly interconnected nodes. A prominent problem is how to investigate the relationship between macro-scale modules that are retrieved by optimizing global network measures, and micro-scale structure that are defined by specific queries of the analysis (e.g., nodal features). By generalizing fundamental concepts of joint space-frequency localization to network theory, here we propose a flexible framework to study interactions between micro- and macro-structure. Similar to pointing and focusing a magnifying glass, the analysis can be directed to specific micro-scale structure, while the degree of interaction with the macro-scale community structure can be seamlessly controlled. In addition, the method is computationally efficient as a result of the underlying low-dimensional optimization problem. 
\end{abstract}

\begin{document}

\flushbottom
\maketitle
\thispagestyle{empty}

\section*{Introduction}
Many scientific disciplines increasingly rely upon network modeling. Insights into basic network organization are often driven by clustering and modularity; i.e., grouping of nodes derived from their within and between cluster connectivity. To reveal such structure, spectral methods constitute an important class of approaches based on the eigendecomposition of graph matrices such a the graph Laplacian or modularity matrix. From an optimization viewpoint, they provide elegant solutions to the convex relaxation of different clustering formulations. However, the eigenvectors that serve for such solutions are global in nature, which means they are localized in the spectral domain, but not in the original graph domain, in accordance with Heisenberg uncertainty principle. 

We provide a fundamental contribution to network theory by introducing the notion of localization, which will identify micro-scale structure, to spectral graph analysis. Our proposal is based on a generalization to graphs of the concept of Slepian functions that form a basis of band-limited functions on classical Cartesian domains with optimal energy concentration in a pre-defined interval~\cite{Slepian.1961,Slepian.1978}. These functions allow exploring the trade-off of joint spatio-spectral localization as both the bandwidth and the interval can be chosen. Slepian functions have been adapted to the sphere in the context of geophysics~\cite{Simons.2006}. Here we developed a new  framework that generalizes Slepian functions to graphs and has the ability to analyze interactions between local and global graph structure; e.g., how selected nodes interact with macro-scale communities. We start by elaborating the framework when the  spectrum is derived from the graph Laplacian, which allows establishing the link with harmonic analysis and classical Slepian functions. Then, we extend the framework for the modularity spectrum, which leads to useful applications for network analysis.

\section*{Results}
\begin{figure*}
  \centering
  \includegraphics[width=15cm]{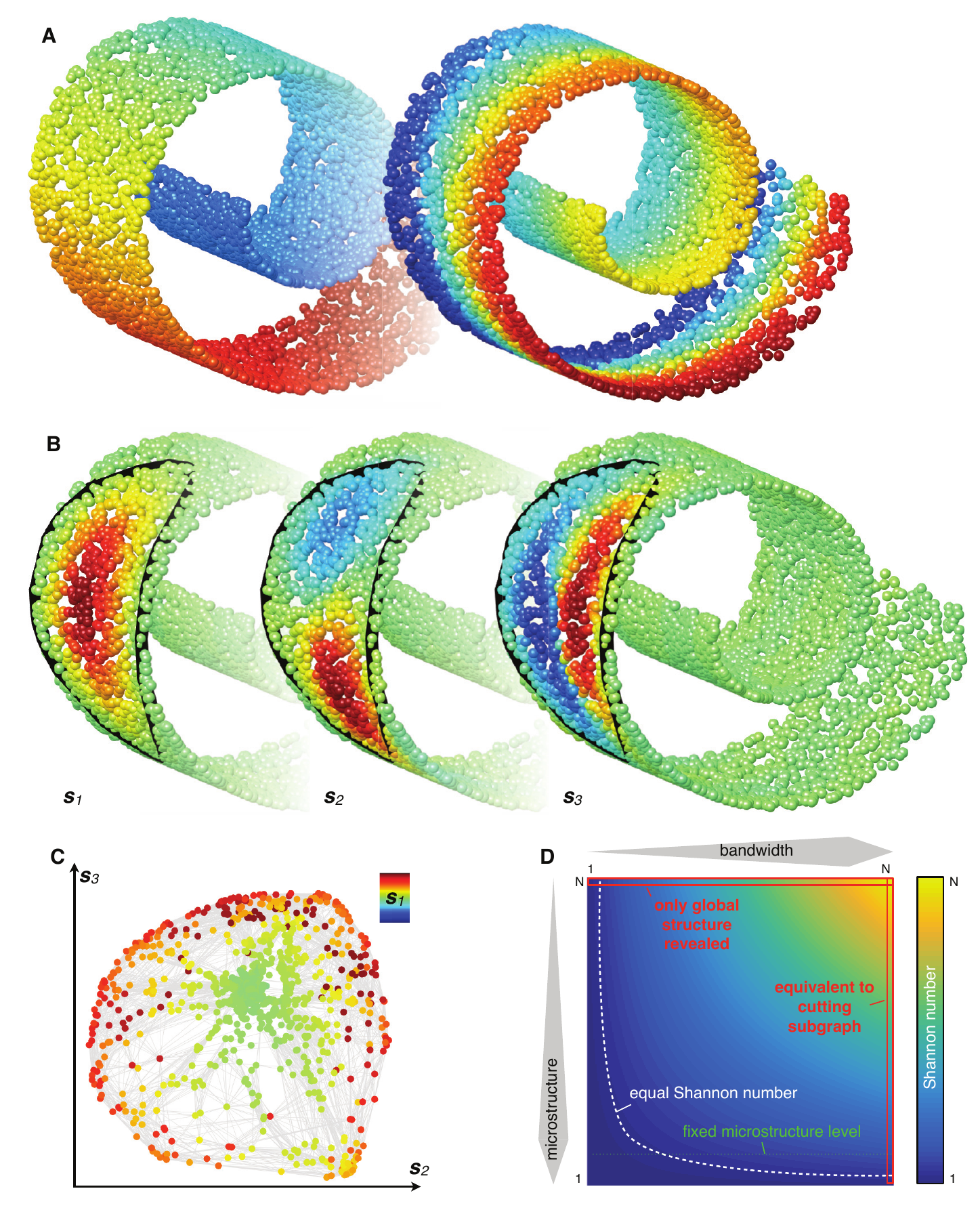}
  \caption{\label{fig:swissroll} Swiss Roll dataset example for 4'400 nodes. (A)~First eigenvectors of the graph Laplacian. (B)~First Slepian vectors $\vc s_k$ for the subgraph that contains the nodes inside the black contour. (C)~Combining information of the first three Slepian vectors in a single view. (D)~Shannon number reflects trade-off between bandwidth and microstructure (i.e., size of the subgraph).}
\end{figure*}

Networks are characterized by the adjacency matrix $\ma A$, where the elements $A_{i,j}$ indicate the weight of the connection between nodes $i,j=1,\ldots,N$. The graph Laplacian is defined as $\ma L=\ma D-\ma A$, where $\ma D$ is the diagonal degree matrix with elements $D_{i,i}=\sum_j A_{i,j}$. We assume undirected graphs for which $\ma{A}$ is symmetric and the graph Laplacian can be defined with different normalizations~\cite{Chung.1997}. The proposed framework can be applied to any form of the Laplacian that is most relevant for the application at hand. The eigenvectors $\vc u_k$ of $\ma L$ satisfy $\ma L\vc u_k=\lambda_k \vc u_k$ with eigenvalues $\lambda_1=0\le \lambda_2 \le \ldots \le \lambda_N$; the eigenvectors play the role of basis vectors of the graph spectrum, and the associated eigenvalues of frequencies~\cite{Chung.1997}. As an illustration, in Fig.~\ref{fig:swissroll}-A, the first eigenvectors of the graph Laplacian of the ``Swiss roll' dataset are represented. These graph signals correspond to our intuition about slow variations along the main axes of the stretched and rolled ribbon on which the nodes are defined.

To generalize Slepian functions to graphs, first, we introduce the notion of ``bandlimiting'' the graph spectrum; i.e., we restrict the spectrum to the eigenvectors with the $W<N$ smallest eigenvalues and group them in the matrix $\ma U_W$ of size $N\times W$. Recent work~\cite{Agaskar.2013} has demonstrated that the span of these first $W$ eigenvectors achieves a lower bound for the product of signals spread in the graph and the spectral domains. Second, we define a subgraph $\mathcal{S}$ that contains the $S$ nodes in which we want the energy concentration to be maximal. The selection of the nodes in $\mathcal{S}$ can be represented by a diagonal matrix $\ma{S}$ where the elements $S_{i,i}=0/1$ indicate the presence of a node $i$ in $\mathcal{S}$. Finding the linear combination of eigenvectors $\ma{U}_W \vc{v}$ within the bandlimit $W$ and with maximal energy in $\mathcal{S}$ reverts to optimizing the Rayleigh quotient $\vc v^T \ma C \vc v / \vc v^T \vc v$,
where $\ma{C}=\ma{U}_W^T \ma{S} \ma{U}_W$ is the concentration matrix. 
Finding the optimal vector $\vc v$ translates into an eigendecomposition problem, $\ma C\vc v_k=\mu_k \vc v_k$, where $\mu_k$ represents the energy concentration of the Slepian vector $\vc s_k=\ma U_W\vc v_k$ in $\mathcal{S}$. The Slepian vectors are orthonormal over the whole graph as well as orthogonal over the subgraph $\mathcal{S}$ (See Methods). By convention, they are sorted according to decreasing energy concentration $1>\mu_1\ge \mu_2 \ge \ldots > 0$ in $\mathcal{S}$. 
We illustrate the procedure on the ``Swiss roll'' in Fig.~\ref{fig:swissroll}-B. The subgraph $\mathcal{S}$ contains the nodes within the black contour. The bandwidth ratio $W/N$ is limited to 1\%. The first three Slepian vectors are highly concentrated in the subgraph and build up a localized spectral decomposition; i.e., the underlying Laplace eigenvectors $\ma{U}_W$ pass on their spectral properties to the Slepian vectors. The embedding in Fig.~\ref{fig:swissroll}-C plots all nodes according to the second and third Slepian vectors and colors them according to the first one. In this view, nodes well inside the subgraph $\mathcal{S}$ are represented at the outside of the embedding and organized according to orientation, while nodes not part of the subgraph are pulled towards the center. The Shannon number $K=WS/N$ indicates the number of Slepian vectors that are strongly concentrated in $\mathcal{S}$. This trade-off is illustrated in Fig.~\ref{fig:swissroll}-D and an example of the effect on the first Slepian vector in Fig.~\ref{fig:swissroll-spectrum}. For more irregular networks than the Swiss roll, the Shannon number for a fixed subgraph $\mathcal{S}$ becomes dependent on the topology of the network, but it can be derived empirically from the phase-transition that occurs in the sequence $\mu_k$, $k=1,\ldots,W$. The interesting regime for the bandwidth is where the transition between macro-scale and micro-scale organization occurs and interactions can be made visible. 

Spectral graph theory relates to the classical graph cut problem that consists of partitioning the graph into clusters of nodes such that the cut size is minimized; i.e., the total weight of the edges that run between clusters. Depending on the chosen normalization of the Laplacian, the definition of the cut size varies with respect to the original network specification. For two clusters indicated by $g_i=\pm 1$, the cut size is defined as $R=\frac{1}{4} \sum_{i,j} A_{i,j} (1- g_i g_j)=\frac{1}{4} \sum_{i,j} L_{i,j} g_i g_j$. 
Finding clusters that minimize $R$ is NP hard, but can be relaxed by allowing $g_i$ to take any real value, which can then be solved by taking the eigenvector of the Laplacian with the second-smallest eigenvalue, the Fiedler vector~\cite{Fiedler.1989}. The Fiedler vector can be thresholded, but, in many practical cases, additional vectors are taken into account for multi-class clustering and to enable discovery of richer network structure~\cite{Ng.2002}. The first eigenvectors can also be used as an effective way to reduce dimensionality and visualize the network in a low-dimensional space with topological relevance. 
From this perspective, the Slepian vectors solve the graph cut problem using the $W$ most important eigenvectors with maximal energy concentration in the subgraph $\mathcal{S}$. 
Computationally, the first $W$ eigenvectors of the Laplacian can be obtained using efficient iterative numerical procedures~\cite{Lehoucq.1996}, followed by the determination of the Slepian vectors that requires the solution of only a low-dimensional eigendecomposition (i.e., the concentration matrix $\ma{C}$ is only of size $W\times W$). 

\begin{figure*}
  \centering
  \includegraphics[width=15cm]{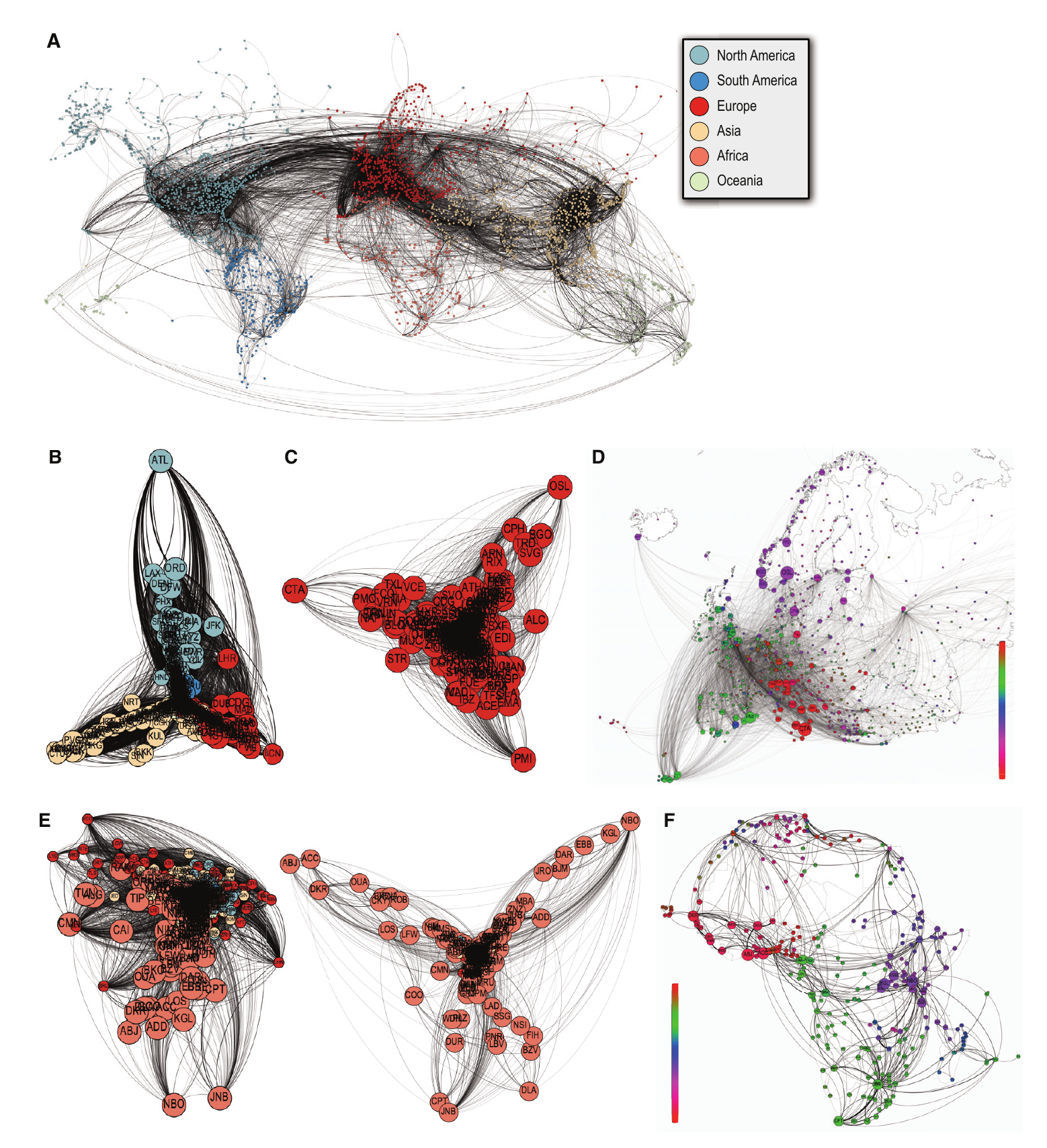}
  \caption{\label{fig:openflights} (A)~OpenFlights Airports Database visualization. (B)~Global community structure. Nodes are scaled according to their degree and labeled with their IATA code. Interactions between macro-scale community structure and selected subgraphs: (C)~European airports with their (D)~geographical embedding; (E)~African airports for low- and high bandwidth and their (F)~geographical embedding. Selected nodes are enlarged.}
\end{figure*}

This insight permits to extend the graph Slepian framework to other spectral approaches, in particular, for modularity analysis~\cite{Newman.2006}. Communities are clusters of nodes that are more strongly interconnected between them than with other communities. For two communities indicated by $g_i=\pm 1$, the main concept is to define the modularity function $Q=\sum_{i,j} (A_{i,j} - P_{i,j}) g_i g_j$, where $P_{i,j}=D_{i,i} D_{j,j} / \sum_{i,j} D_{i,j}$ represents the expected weight of an edge between nodes $i$ and $j$~\cite{Girvan.2002}. The spectral method for community detection then forms the modularity matrix $\ma B=\ma A-\ma P$ and solves the eigendecomposition $\ma B\vc u_k=\lambda_k \vc u_k$. Community structure can be derived from the eigenvectors with largest positive eigenvalues. Similar as for the graph Laplacian, the modularity spectrum can be bandlimited by considering $W<N$ eigenvectors 
and building the concentration matrix $\ma C$ for a given subgraph. We analyzed data taken from the OpenFlights Airports Database with 3'281 airports and 67'202 routes, illustrated in Fig.~\ref{fig:openflights}-A where airports are colored according to continents. In Fig.~\ref{fig:openflights}-B, community structure revealed by the first two eigenvectors of $\ma{B}$ shows three different branches along which airports in North America, Europe, and Asia are arranged, respectively. The airports that are most densely connected within their community are at the end of the branches; e.g., Atlanta (ATL) and Barcelona (BCN). Airports that serve as the largest connector hubs for inter-continental flights stand out between the branches; e.g., London Heathrow (LHR) and New York (JFK) between North America and Europe, Tokyo (NRT) and Honolulu (HNL) between North America and Asia, and Singapore (SIN), Bangkok (BKK), Kuala Lumpur (KUL) between Asia and Europe. The global community structure of airline routes can be further explored by taken into account additional eigenvectors (see Fig.~\ref{fig:global}), but it is mostly explained by continents, which is consistent with previous reports~\cite{Guimera.2005}. 


We then wanted to investigate more subtle interactions between selected airports and this macro-scale community structure. We defined the subgraph $\mathcal{S}$ containing all European airports ($662$ nodes) and embed the nodes in a new topological space defined by the coordinates of the two first Slepian vectors. The bandwidth $W$ provides a tuning parameter for the trade-off between local (within the subgraph) and global community structure. Specifically, for low bandwidth, this view is identical to the global community structure from Fig.~\ref{fig:openflights}-B. For high bandwidth, the European branch opens up and the intra-European structure is revealed. For instance, in Fig.~\ref{fig:openflights}-C, we identify at bandwidth $W=80$ three major branches along which North, West, and East European airports are organized, respectively. Similar to the interpretation of Fig.~\ref{fig:swissroll}-C, the nodes that are farthest from the center are most interior to the communities. 
To better illustrate this observation, in Fig.~\ref{fig:openflights}-D, the nodes are mapped to their geographical position and the angle and magnitude with respect to the center in the Slepian view is used to determine their color and size, respectively. 

Next, we wanted to study interactions of the African airports ($288$ nodes), which remain invisible in the global view of Fig.~\ref{fig:openflights}-B. In Fig.~\ref{fig:openflights}-E, we first show a low-bandwidth view ($W=20$) where mainly two branches of African airports form. Interestingly, presence of some non-African airports is prominent; e.g., Paris (CDG) with many routes to airports in the Maghreb including Casa Blanca (CMN), Tunis (TUN), and Algiers (ALG); and Brussels (BRU) with historical routes to West Africa including Abidjan (ABJ), and Dakar (DKR), Ouagadougou (OUA). When increasing the bandwidth ($W=80$), three branches can be distinguished that further confirm modular organization of the African routes within West Africa, Central and South Africa (e.g., Johannesburg (JNB)), and East Africa (e.g., Nairobi (NBO))---see Fig.~\ref{fig:openflights}-F for the geographical embedding that confirms this interpretation. The organization of the North African airports is less pronounced because they interconnect much denser to the European continent and, therefore, take a more central position in the Slepian view. In Fig.~\ref{fig:gradual}, the effect of progressively increasing the bandwidth is illustrated in detail. 

\begin{figure*}
  \centering
    \includegraphics[width=14cm]{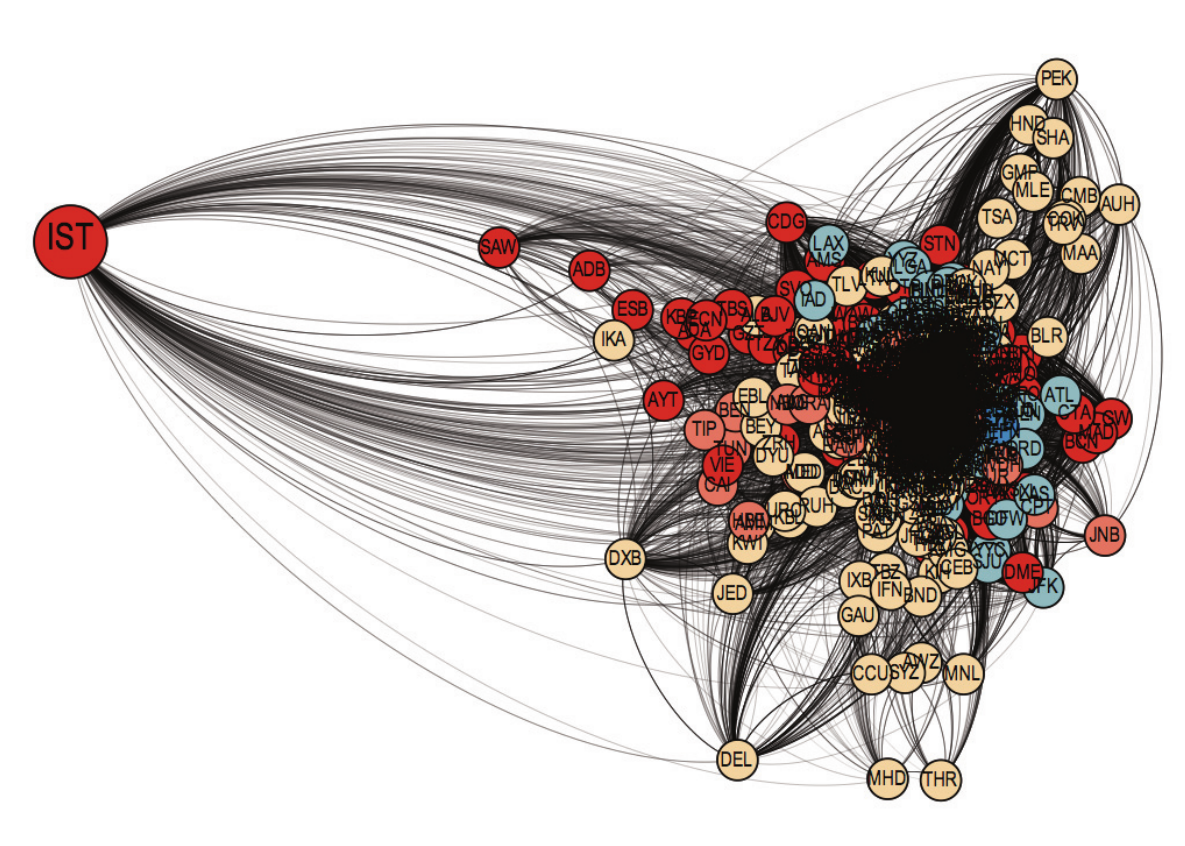} 
  \caption{\label{fig:IST} Interactions between subgraph $\mathcal{S}$ that contains a single node (i.e., Istanbul Atat\"urk (IST)) and the global community structure. Bandwidth has been chosen at $W=30$.}
\end{figure*}

The framework can be pushed to its extreme by including a single node in the subgraph. In Fig.~\ref{fig:IST}, the embedding provided by the first two Slepian vectors for the subgraph $\mathcal{S}$ that only includes a single node (i.e., Istanbul Atat\"urk (IST)) is shown. Even for relatively low bandwidth ($W=30$), the result in Fig.~\ref{fig:IST} shows how other Turkish airports such as Ankara (ESB) and ADB (Izmir) interact closely with IST, but also Tehran (IKA) for which IST is one of the main gateways. Interestingly, Istanbul's other international airport, Sabiha G\"ok\c cen (SAW), is topologically closest to IST despite the lack of a direct route between IST and SAW.

\section*{Discussion}
In sum, the proposed framework provides a novel way to put a magnifying glass on network organization and study interactions between micro- and macro-structure. The selection of the micro-structure can be compared to where to direct the magnifying glass, while tuning the bandwidth to the degree of focus on the selection. Due to the underlying low-dimensional optimization problem, the method is computationally efficient and  allows for a versatile and interactive exploration of the network properties. 
We expect this framework to serve further methodological developments in this active field~\cite{Shuman.2013,Chen.2015}. For instance, the discovery of hierarchical structure~\cite{Ravasz.2002,Arenas.2008,Expert.2011,Irion.2014,Hero.2015} could benefit from a gradual refinement of the subgraph definition while using the same underlying global eigenvectors. Spectral approaches also play a central role in recent multi-resolution approaches~\cite{Hammond.2011,leonardi1302,Tremblay.2014}, for which the Slepian vectors could become localized basis functions to design graph wavelets. Finally, the notion of the subgraph selection could also be extended to include layers, as in multiplex networks~\cite{Mucha.2010}, and to be specified in terms of links instead of nodes, as in link communities~\cite{Palla.2005,Ahn.2010}.

\section*{Methods}
The core of the framework consists of solving the eigendecomposition of the concentration matrix $\ma{C}=\ma{U}_W^T \ma{S} \ma{U}_W$, where $\ma{S}$ is the selection matrix of the subgraph $\mathcal{S}$ and $\ma{U}_W$ contains the $W$ first smallest or largest eigenvectors of the graph Laplacian or modularity matrix, respectively. This leads to the eigenvectors and eigenvalues that satisfy $\ma{C}\vc{v}_k=\mu_k \vc{v}_k$, $k=1,\ldots,W$. The eigenvalues $\mu_k$ represent the energy concentration in $\mathcal{S}$ of the Slepian vectors $\vc{s}_k=\ma{U}_W \vc{v}_k$ that are a linear combination of the original eigenvectors $\vc{u}_k$, $k=1,\ldots,W$. By convention, the Slepian vectors are sorted according to decreasing energy concentration as $1> \mu_1 \ge \mu_2 \ge \ldots \ge \mu_W > 0$. The sequence of eigenvalues $\mu_k$ presents a phase transition between well-localized and poorly-localized eigenvectors. In the classical case, this phase transition occurs at the Shannon number $K=WS/N$. We observed that $K$ still provides a good rule-of-thumb for the number of Slepian vectors with high energy concentration, but in general it can be derived empirically from the sequence $\mu_k$ itself.

The Slepian vectors $\vc{s}_k$, $k=1,\ldots, W$, satisfy various remarkable properties similar to their classical counterparts. For instance, since they are derived using two eigendecompositions in cascade, they are orthonormal over the whole network, as well as orthogonal over the subgraph $\mathcal{S}$. This can be readily shown as
\begin{eqnarray}
  \vc{s}_k^T \vc{s}_l & = & \vc{v}_k^T \ma{U}_W^T \ma{U}_W \vc{v}_l = \vc{v}_k^T \vc{v}_l = \delta_{k-l}, \\
  \vc{s}_k^T \ma{S} \vc{s}_l & = & \vc{v}_k^T \underbrace{\ma{U}_W^T \ma{S} \ma{U}_W}_{=\ma{C}} \vc{v}_l = \vc{v}_k^T \ma{C} \vc{v}_l = \mu_l \delta_{k-l},
\end{eqnarray}
where $\delta$ is the Kronecker delta. 


\subsection*{Swiss Roll}
The ``Swiss roll'' dataset was generated using the code available from~\cite{Harmeling.2007} that embeds nodes with randomly generated positions (uniformly in a unit square) into a ribbon that was stretched and rolled in three dimensions. The number of nodes was set to $N=4'400$. The Euclidean distances $d_{i,j}$ between all pairs of nodes $(i,j)$ were computed and converted into a weighted adjacency matrix by the kernel operation $A_{i,j} = \exp(-d_{i,j}^2 / 0.005)$. Weights smaller than 0.01 were put to zero (which corresponds to distances larger than $0.15$). To determine the subgraph $\mathcal{S}$, we selected the node that was most to the left and in the middle of the roll. We selected all neighboring nodes within a distance $0.8$, which corresponded to $762$ nodes. We then determined the normalized graph Laplacian $\ma{L}=\ma{D}^{-1/2} (\ma{D}-\ma{A}) \ma{D}^{-1/2}$. 

\begin{figure*}
  \centering
    \includegraphics[width=16cm]{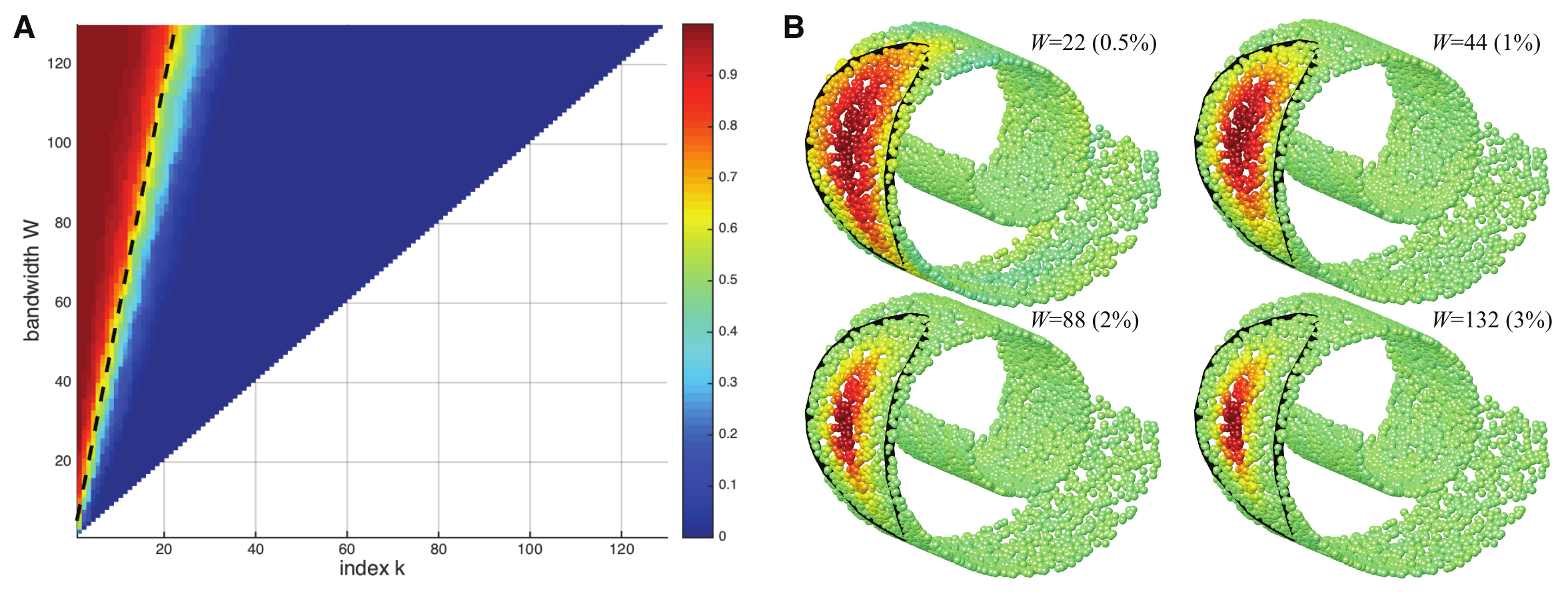} 
  \caption{\label{fig:swissroll-spectrum} (A)~Sequence of the eigenvalues $\mu_k$, $k=1,\ldots,W$ for bandwidth $W=2,\ldots,130$. The dashed line indicates the Shannon number $K$. (B)~First Slepian vector for different choices of the bandwidth.}
\end{figure*}

The sequence of eigenvalues $\mu_k$, $k=1,\ldots,W$, is shown in Fig.~\ref{fig:swissroll-spectrum}-A for W varying between $2$ and $130$. The dashed line indicates the Shannon number $K$ that provides a good estimate of the phase transition in $\mu_k$. The computation time (Apple MacBook Pro, 2.6GHz i7/16GB RAM, running Matlab R2014b) was 0.4 seconds to obtain the $130$ eigenvectors $\ma{U}_W$ of $\ma{L}$ with smallest eigenvalues. The second eigendecomposition to retrieve the Slepian function took less than 0.02 seconds. 

In Fig.~\ref{fig:swissroll-spectrum}-B, the first Slepian vector is visualized in its 3-D space for increasing bandwidth. For higher bandwidth, the energy concentration in the selected subgraph becomes stronger. For the illustration in the main text, we set the bandwidth to $W=44$ or $1\%$ in relative terms. 

\subsection*{OpenFlights Airports Database}
We imported the OpenFlights Airports Database from May 20, 2015. Airports without any routes were removed; which resulted in $3'281$ airports. The undirected weighted adjacency matrix was built as the number of routes between each pair of airports independent of the direction. In total, $67'202$ routes were taken into account. The adjacency matrix is stored as a sparse matrix with $38'047$ non-zero entries. We used the modularity matrix $\ma{B}$ with the Girvan-Newman null model. This matrix is evaluated by an inline-function to take full benefit of its sparse structure.  The computation of the $W$ eigenvectors $\ma{U}_W$ of $\ma{B}$ with largest eigenvalues took 0.5 seconds. The second eigendecomposition to retrieve the Slepian vectors took less than 0.01 seconds for all subgraphs and maximal bandwidth ($W=100$). 

\begin{figure*}
  \centering
    \includegraphics[width=15cm]{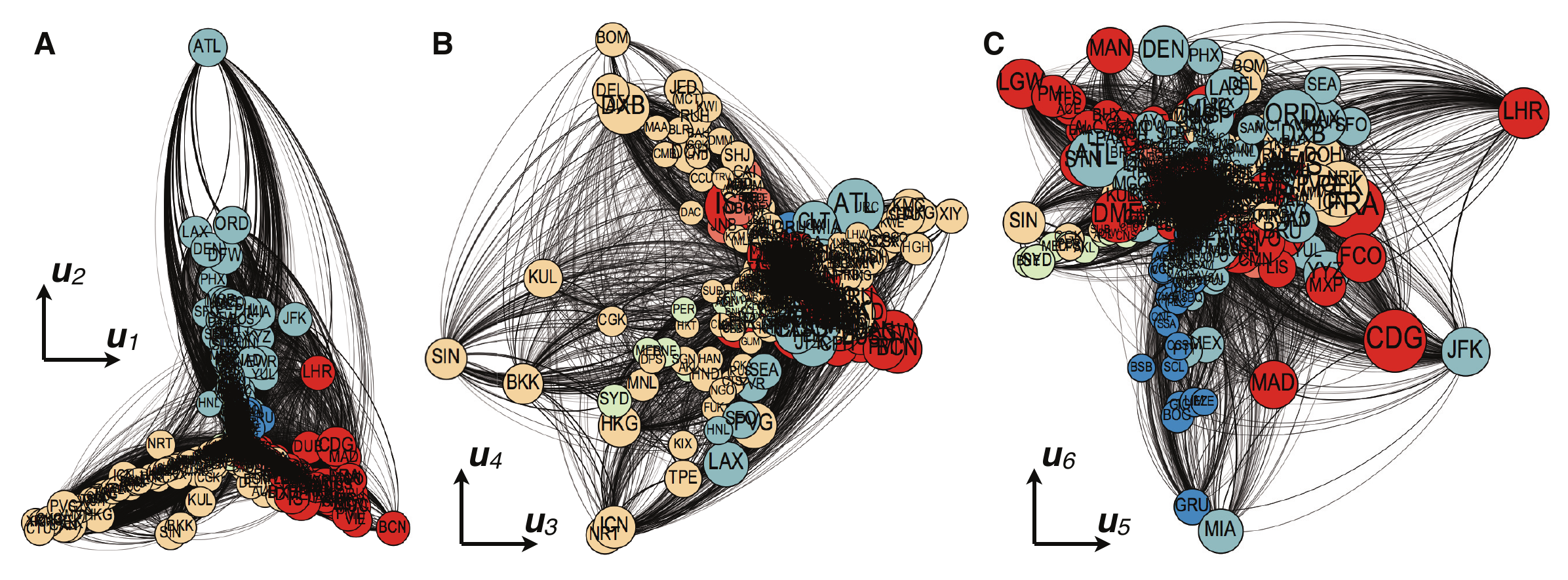} 
  \caption{\label{fig:global} Global community structure revealed by the first 6 eigenvectors of the modularity matrix. Nodes are scaled according to their degree and labeled with their IATA code.}
\end{figure*}

\begin{figure*}
  \centering
    \includegraphics[width=16cm]{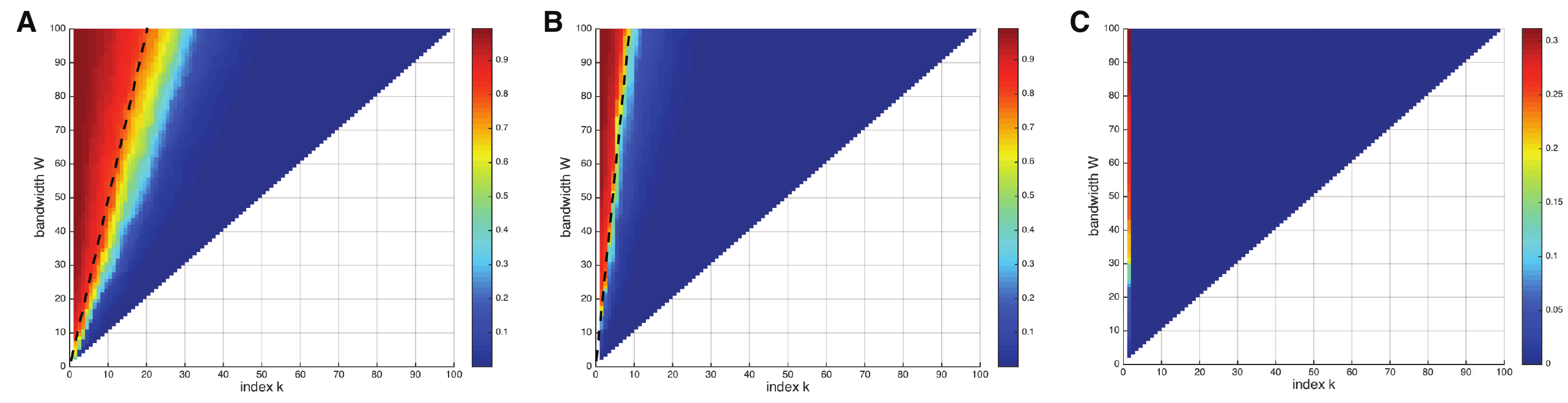} 
  \caption{\label{fig:spectra} Sequence of the eigenvalues $\mu_k$, $k=1,\ldots,W$ for various choices of the subgraph: (A)~European airports; (B)~African airports; (C)~Istanbul Atat\"urk airport (IST).}
\end{figure*}

In Fig.~\ref{fig:global}, we show the macro-scale community structure revealed by the first 6 eigenvectors of the modularity matrix. The first view given by $(\vc{u}_1,\vc{u}_2)$ shows three branches along which airports in the American, European, and Asian continent are arranged. The second and third views by $(\vc{u}_3,\vc{u}_4)$ and $(\vc{u}_5,\vc{u}_6)$ show additional refinements to this organization, but they become increasingly difficult to interpret visually, mainly due to the orthogonality constraints of the eigenvectors $\vc{u}_k$. The Slepian framework will use $W$ eigenvectors to `refocus' onto a predefined subgraph $\mathcal{S}$.

In Fig.~\ref{fig:spectra}, we show the Slepian spectrum (i.e., sequence of eigenvalues $\mu_k$, $k=1,\ldots, W$) depending on $W$ for various choices of the subgraph $\mathcal{S}$: Europe (662 nodes), Africa (288 nodes), Istanbul Atat\"urk (1 node). The black dashed line indicates the Shannon number, which still provides a good rule of thumb despite the more irregular structure of the network. 

\begin{figure*}
  \centering
    \includegraphics[width=16cm]{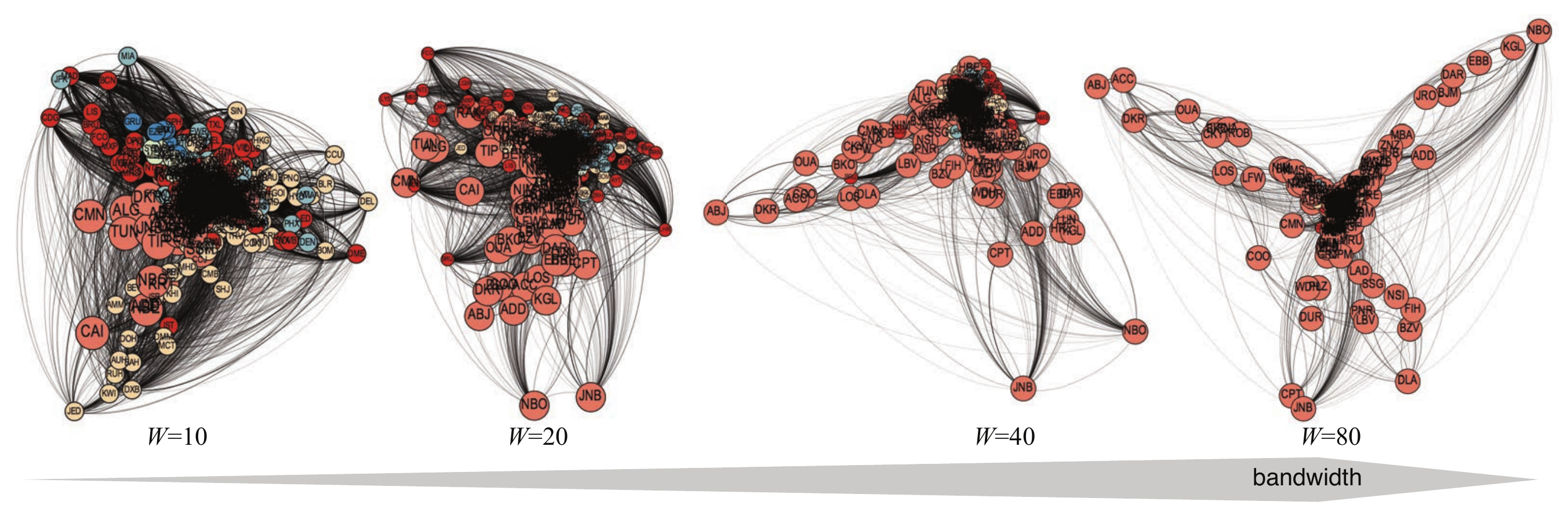} 
  \caption{\label{fig:gradual} Interactions between subgraph containing African airports and the macro-scale community structure for gradually increasing bandwidth: $W=10, 20, 40, 80$.}
\end{figure*}

In Fig.~\ref{fig:gradual}, we show the visualization based on the first two Slepian vectors when the subgraph contains all African airports and the bandwidth $W$ is gradually increased from $10$ to $80$. For $W=10$, a branch with African airports appears while the European and Asian branches from the macro-scale view are still present. For $W=20$, more diversity in the intra-African organization is present, including European airports that play a central role (e.g., Brussels (BRU)). For $W=40$ and $W=80$ the within-Africa community structure is further confirmed by several branches and the central hubs of each community taking a place at the ends of these branches.  

Upon acceptance of this manuscript, the source code (Matlab) and experimental data will be made available to the community. In addition, we are developing a Gephi plugin (Java) for interactive visualization.


\section*{Additional information}
The author declares no competing financial interests.

\end{document}